\documentclass[article,aps,prd,12pt,notitlepage]{revtex4}
\usepackage{amsmath,graphicx}
\usepackage{bm}
\usepackage{amssymb}
\usepackage{graphicx,xcolor}
\usepackage{epigraph}
\usepackage{csquotes}
\usepackage{amsmath,amssymb,stmaryrd}
\begin{document}
\def\e{\enquote}
\def\tr{\rm{Tr}}
\def\la{{\langle}}
\def\ra{{\rangle}}
\def\a{{\alpha}}
\def\q{\quad}
\def\ta{t_0}
\def\w{\tilde}
\def\om{\omega}
\def\t{\tilde{t}}
\def\a{\hat{A}}
\def\H{\mathcal{H}}
\def\N{\mathcal{N}}
\def\h{\hat{H}}
\def\E{\mathcal{E}}\def\la{{\langle}}
\def\u{\hat U}
\def\U{\hat U}
\def\C{\hat C}
\def\D{Q}
\def\S{\tilde S}
\def\A{{\mathcal A}}
\def\AA{{\tilde A}}
\def\AAA{{\hat A}}
\def\B{{\hat  B}}
\def\Delt{\tilde \Delta}
\def\QQ{\hat S}
\def\ppi{\hat \pi}
\def\R{\text {Re}}
\def\I{\text {Im}}
\def\e{\enquote}
\def\qq{s}
\def\up{\uparrow}
\def\do{\downarrow}
\def\Q{\hat Q}
\def\fb{\overline F}
\def\wb{\overline W}
\def\nl{\newline}
\def\h{\hat H}
\def\ff{\overline q}
\def\k{\overline k}
\def\F {Q}
\def\f{q}
\def\lm{\lambda}
\def\lmu{\underline\lambda}
\def\q{\quad}
\def\t{\tau}
\def\f{\overline f}
\def\l{\ell}
\def\n{\\ \nonumber}
\def\ra{{\rangle}}
\def\Ep{{\mathcal{E}}}
\def\T{T_{total}}
\def\M{{\mathcal{M}}}
\def\omga{{\epsilon}}
\def\+{\text{up}}
\def\-{\text{down}}
\def\h{\hat{H}}
\def\b{\color{blue}}
\def\r{\color{black}}
\def\g{\color{green}}
\title{Speakable and unspeakable in quantum measurements}
\author{D. Sokolovski$^{a,b,c}$}
\email{dgsokol15@gmail.com}
\author{D. Alonso$^{d}$}
\author{S. Brouard$^{d}$}
\affiliation{$^{a}$ Departmento de Qu\'imica-F\'isica, Universidad del Pa\' is Vasco, UPV/EHU, 48940, Leioa, Spain}
\affiliation{$^{b}$ IKERBASQUE, Basque Foundation for Science, Plaza Euskadi 5, 48009, Bilbao, Spain}
\affiliation{$^{c}$ EHU Quantum Center, Universidad del Pa\' is Vasco, UPV/EHU, 48940, Leioa, Spain}
\affiliation{$^{d}$ Departamento de F\'isica y IUdEA, Universidad de La Laguna,La Laguna,Tenerife, Spain}
 \date\today
%
%
\begin{abstract}
\begin{center}
{\bf ABSTRACT}
\end{center}
\noindent
{Quantum mechanics, in its orthodox version, imposes severe limits on what can be known, or even said, about the 
condition of a quantum system between two observations.
A relatively new approach, based on so-called \e{weak measurements}, suggests that such forbidden knowledge can be gained
by studying the system's response to an inaccurate weakly perturbing measuring device. It goes further to propose 
 revising the whole concept of physics variables, and offers various examples of counterintuitive quantum behaviour. 
Both views go to the very heart of quantum theory, and yet are rarely compared directly.
A new technique must either transcend the orthodox limits, or just prove that these limits
 are indeed necessary. 
We study both possibilities,  and
 find for the orthodoxy. }
\end{abstract}

\maketitle

\newpage
\setlength{\epigraphwidth}{0.8\textwidth}
\begin{tiny}
\epigraph{
But when you have no apparatus to determine through which hole the thing goes, 
then you cannot say that it either goes through one hole or the other.
You can always {\it say} it - provided you stop thinking immediately and make no deductions from it.
Physicists prefer not to say it, rather than stop thinking at the moment.}
{R.P. Feynman in {\it Character of Physical Law}}
\epigraph{...we can separate, for example, the spin from the charge of an electron, or internal energy of an atom from the atom itself.}
{Y. Aharonov, S. Popescu,  D. Rohrlich, and P. Skrzypczyk in {\it Quantum Cheshire Cats}}
\end{tiny}
\noindent
\section{Introduction}
{ In 1963 \cite{FeynL} ,1964 \cite{FeynC}, and 1965  \cite{FeynH} Feynman and co-authors outlined what they considered the most general principles
of the quantum method.
At  about the same time, the authors of \cite{ABL} proposed what is known as the \e{Aharonov,  Bergmann, and Lebowitz rule}
for calculating the probabilities of a pre- and post-selected quantum system. 
Feynman's arguments  are rarely revisited in a discussion of the quantum measurement theory.
On the other hand, a recent development of the views expressed in  \cite{ABL} has given the reader the concept of \e{weak measurements}, 
and a plethora of \e{quantum paradoxes} the concept helps to establish (see, for example, \cite{SPIN100}-\cite{CAT2}). 
\newline
The attitude to asking questions about the quantum method is often summed by Mermin's taunt \e{Shut up and calculate} \cite{MERM}.
Calculations are, no doubt, useful, but understanding the method's  limitations, could tell one something about 
the relation between a human observer and nature. 
\newline
The purpose of this paper is to compare directly the orthodoxy of \cite{FeynL}  (one can hardly be more orthodox than when appealing to a reputable undergraduate text)
with recent developments in the quantum measurement theory  \cite{SPIN100}-\cite{CAT2}, \cite{REVIEW1}-\cite{MES4}. Our aim is to see whether anything new has indeed 
been added to the views expressed in \cite{FeynL} - \cite{FeynH}, or if at least  some of these views have been disproved by the more recent research (a preliminary
attempt of such a comparison was made in \cite{DSmyst}). It is easy to see   \cite{DSABL} that the original ABL rule is a particular 
example of the Feynman's rule for assigning probabilities \cite{FeynL}. A further comparison is more involved. 
Feynman's reasoning is subtle, as  one has to walk a \e{logical tightrope} in order to \e{interpret nature} \cite{FeynC}.
{\r The centrepiece of the orthodox description is the Uncertainty Principle, best known in its Heisenberg form  \cite{FeynL}.
According to the Principle, certain questions, easily answered in a classical context, should have no answer when dealing with a quantum  system.}
In particular, very little may be said about the path taken by an unobserved  particle in a double-slit experiment,  \cite{FeynL},  \cite{FeynC} 
and, by the same token, about the  state of any system between two observations. 
The authors of \cite{COMPL} suggested that a meaningful description of a system between the measurements can be obtained 
by employing weakly perturbing detectors.
{\r (We refer the reader to Ref.\cite{Lund} for a possible use of such detectors in order to surpass the Heisenberg limit in joint measurements
of a particle's position and momentum.)
Over the years, 
the approach proposed in \cite{COMPL} has lead to a number of spectacular claims \cite{COMPL}-\cite{CAT2}, which we will submit to further scrutiny here.}
\newline
The rest of the paper is organised as follows.
In Sect. II we discuss consecutive quantum measurements, the need to produce records of the past outcomes, 
and a simplified description, possible if particular  types of probes are used.
Sect. III reviews the \e{orthodox} description of quantum behaviour, given in \cite{FeynL}, \cite{FeynC}, \cite{FeynH}. 
In Sect. IV we use the description to analyse a quantum measurement analogy of the double-slit problem. 
In Sect. V we leave the narrative of \cite{FeynL}, \cite{FeynC}, \cite{FeynH}, and discuss the \e{weak values}, first
introduced in \cite{SPIN100}, as well as their possible interpretations. 
In Sect. VI we introduce a three-slit (\e{three-box}) system, and create our own \e{paradox}, similar to that proposed 
in \cite{3BOX}. 
In Sect. VII the \e{paradox} is dismissed for reasons explained in Sect. V.
In Sect. VIII we revisit a  four-slit (quantum Cheshire cat \cite{CAT1}, \cite{CAT2}) problem.
Sect. IX contains our conclusions. 
Appendices A and  B contain the necessary technical details. 
To condense the presentation, we will bring in the proverbial Alice (an orthodox theorist), Bob (an experimenter), and Carol, a kind of  \e{devil's advocate}, 
who disagrees with Alice and asks her awkward questions. 
We apologise to the reader who may find these characters annoying. 
\section{Bob wants to measure three quantities}
Bob has a quantum system, and wants to measure, at times $t_0 <t_1 < t_2$, the quantities ${\bf A}$, ${\bf B}$, and ${\bf C}$, each taking certain number of
discrete values $M_{A,B,C}\le N$.  He asks Alice the theorist to tell him how to do it, and what he is likely to see.
\newline
Alice tells him that at the end of the experiment, at $t=t_2$, he would need three visible records of the measured values
(the system itself cannot provide this information). He would then measure the frequency with which a particular series of records
occurs after many trials. 
\newline
Alice's mathematical task is to predict the said frequencies, and to do so she associates with the system an $N$-dimensional Hilbert space $\mathcal H^S$, and represents
Bob's quantities by Hermitian operators, $\hat A$, $\B$, and $\C$, acting therein. (If a classical calculation were to be made, she would consider a phase space, and dynamical functions of positions and momenta).
There are three orthonormal eigenbases of  $\hat A$, $\B$, and $\C$, $\left\{|a_k\ra\right\}$, $\left\{|b_l\ra\right\}$, and $\left\{|c_n\ra\right\}$.
The first measured quantity, she tells Bob, must not have degenerate eigenvalues, so that
if the first measurement yields $A_k$, Alice knows that the system is \e{in a state $|a_k\ra$}, and can proceed with 
her calculation. The second operator may have $M_B< N$ degenerate eigenvalues, 
\begin{eqnarray}\label{1}
\hat B=\sum_{m=1}^{M_B} B_m\;\hat\pi(B_m),\q\q 
\hat\pi(B_m)=\sum_{l=1}^N \Delta(B_m-\la b_l|\hat B|b_l\ra) |b_l\ra\la b_l|,
\end{eqnarray}
where $\Delta(X-Y)=1$ if $X=Y$, and $0$ otherwise.
The same applies to the third operator, $\hat C$, but for now we will assume that all $C_n$ are different. 
\newline
Bob's records are to be produced and kept by three probes $D^{A}$, ${D^B}$, and ${D^C}$, each with its own Hilbert space, $\mathcal H^{D^A}$, $\mathcal H^{D^B}$, and $\mathcal H^{D^C}$, so the large Hilbert space which Alice will use for her calculations is a tensor product of the four, 
$\mathcal H^{S+P} =\mathcal H^S\otimes \mathcal H^{D^B}\otimes\mathcal H^{D^B}\otimes\mathcal H^{D^C}$.
\newline
Alice also needs to construct a Hamiltonian $\h^{S+P}(t)$, and a unitary evolution operator,  $\u^{S+P}= \exp [-i \int_{t_0}^{t_2}\h^{S+P}(t')dt']$
 (we use $\hbar =1$) , also acting  
in $\mathcal H^{S+P}$.
The first probe, $D^A$, should briefly couple to the system at $t=t_0$, in such a manner
that any initial product state $|D^A(0)\ra\otimes |\psi^S\ra$ becomes 
\begin{eqnarray}\label{2}
|D^A(0)\ra\otimes |\psi^S\ra\to \sum_{k=1}^N \la a_k|\psi^S\ra |D^A(A_k)\ra\otimes|a_k\ra, \q\q \left(\la D^A(A_{k'})|D^A(A_k)\ra=\delta_{k'k}\right),
\end{eqnarray}
where $|D^A(0)\ra$ is the probe's initial state, and its final condition $|D^A(A_k)\ra$ tells Bob that the outcome of the first measurement was $A_k$. 
The action of the second probe at $t=t_1$ is defined similarly, 
\begin{eqnarray}\label{3}
|D^B(0)\ra\otimes |\psi^S\ra\to \sum_{m=1}^{M_B}  |D^B(B_m)\ra \otimes\hat{\pi}(B_m)|\psi^S\ra,
\end{eqnarray}
whereas the third probe acts at $t=t_2$ according to (\ref{2}). If the system has a non-zero time independent Hamiltonian, $\h^S$, between 
$t_0$ and $t_1$, and $t_1$ and $t_2$
it undergoes its own unitary evolution with an operator $\u^S(t,t') =\exp[-i\h^S(t-t')]$ (see Fig.1a). 
\newline 
It should be mentioned that at the beginning  Bob has no prior knowledge of the 
system's state $|\psi^S\ra$ in Eq.(\ref{2}). So for him the experiment begins when the first probe reads $A_k$.
The necessary statistics can then be collected, e.g., by repeating the measurements on the same system, 
retaining only the cases where the first outcome is $A_k$.
For Alice the initial state of the composite is $|D^C(0)\ra|D^B(0)\ra|D^A(A_k)\ra|a_k\ra$,
and  the probability 
of having the entire  sequence $\{C_n\gets B_m \gets A_k\}$ is given by (see Fig.1a,  the $\otimes$s henceforth omitted)
\begin{eqnarray}\label{4}
&&P(C_n\gets B_m \gets A_k) =\nonumber\\
&&\sum_{j=1}^N
\left|\la D^C(C_n)|\la D^B(B_m)|\la D^A(A_k)|\la \phi^S_j|\u^{S+P}(t_2,t_0)
|D^C(0)\ra|D^B(0)\ra|D^A(A_k)\ra|a_k\ra\right|^2, 
\end{eqnarray}
so that $\sum_{n=1}^N\sum_{m=1}^{M_B}P(C_n\gets B_m \gets A_k)=1$.
The states $|\phi^S_j\ra$ form an arbitrary orthonormal basis, and the choice $|\phi^S_j\ra =|c_j\ra$ simplifies the evaluation due to the appearance of
a Kronecker $\delta_{jn}$ (see Fig.1b).
\newline
One can say that the operator $\u^{S+P}(t_2,t_0)$ describes an unbroken unitary evolution of the composite $\{system + probes\}$
from just after $t_0$ to just after $t_2$. Or one  can simply say that Alice needs to know matrix elements of a unitary operator $\u^{S+P}(t_2,t_0)$ 
between particular states in the Hilbert space of the composite, and not mention the evolution at all \cite{DSreal}. 
\begin{figure}[h]
\includegraphics[angle=0,width=14cm, height= 8cm]{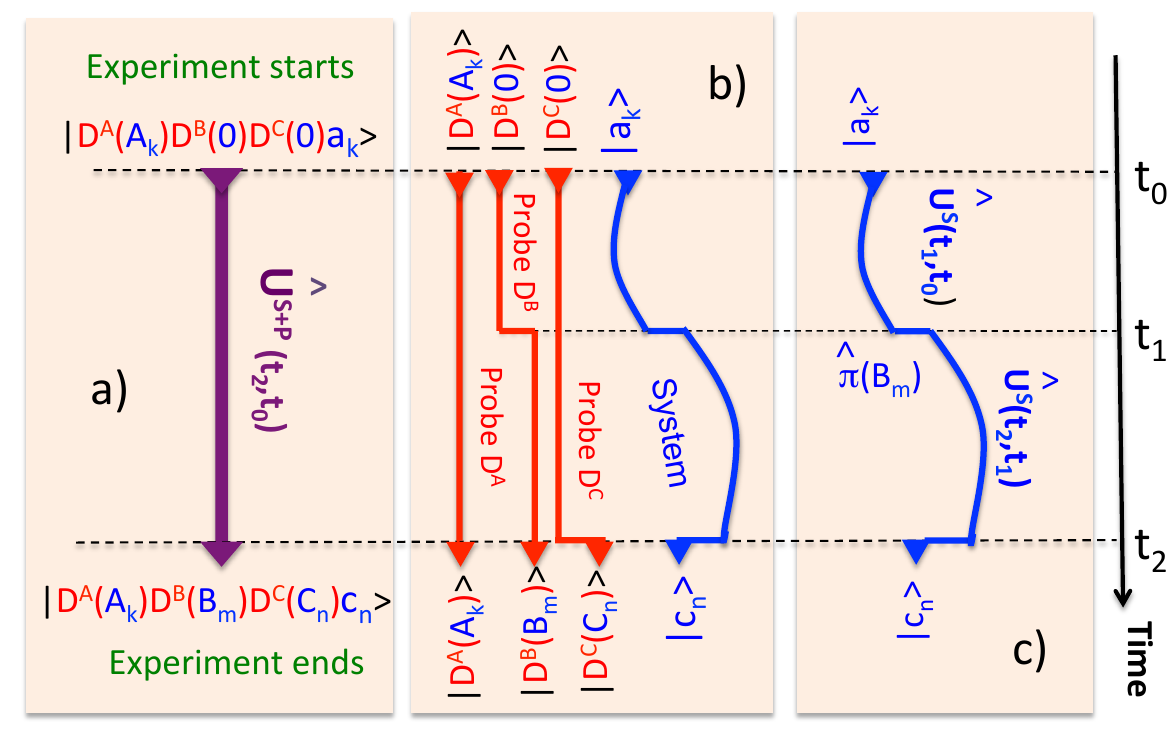}
\caption {a) Unbroken unitary evolution of a composite  $\{system + probes\}$ from $t_0$ to $t_2$ [cf. Eq.(\ref{4})];
b) a more detailed view of the evolution: the probes are coupled at $t_0$, $t_1$, and $t_2$, 
and carry the records until the experiment is finished;  c) broken unitary evolution of the {\it system} only
[cf. Eq.(\ref{2a})].}
\label{Fig.1}
\end{figure}
Furthermore, the probabilities in Eq.(\ref{4}) can be obtained without mentioning the probes explicitly, since 
the couplings (\ref{2}) and (\ref{3}) are simple, and their actions are easily taken into account (details can be found, e.g., in \cite{DSreal}).
In particular, one has for the probability of measuring the sequence ($B_m$, $C_n$), provided the first detection gave $A_k$,
 an equally valid form,
\begin{eqnarray}\label{1a}
P(C_n\gets B_m\gets A_k) =|\la c_n| \u^S(t_2,t_1)\hat\pi(B_m)\u^S(t_1,t_0)|a_k\ra|^2, 
\end{eqnarray}
if the measured eigenvalue of $\C$ is not degenerate. If it is, $\hat C=\sum_{\mu=1}^{M_C} C_{\mu} \hat\pi(C_\mu)$,
an additional summation is needed, 
\begin{eqnarray}\label{2a}
P(C_\mu\gets B_m\gets  A_k) =\sum_{n=1}^N \Delta(C_\mu-\la c_{n}|\hat C| c_{n}\ra)
|\la c_n| \u^S(t_2,t_1)\hat\pi(B_m)\u^S(t_2,t_1)|a_k\ra|^2.
\end{eqnarray}
\newline
Thus, Alice has a choice. She may either consider an unbroken (except at $t=t_2$) unitary evolution 
of the composite $\{system + probes\}$ in a larger Hilbert space $\mathcal H^{S+P}$  [cf. Eq.(\ref{4})],
or she may deal with an evolution in a smaller space $\mathcal H^{S}$, interrupted at $t=t_1$ by insertion of the 
projector $\hat\pi(B_m)$  [cf. Eq.(\ref{2a})]. This interruption is necessary to account for the action of Bob's probe $D^B$ at $t=t_1$, 
 and is by no means mysterious. This second option is clearly more practical.
 
\section{A different view of the same problem: the orthodoxy}
Alice can also reconsider her problem in the light of the general principles of the quantum theory 
as laid out in \cite{FeynL}, \cite{FeynC}.
These principles, we recall, are:
\begin{itemize}
\item[(a)] The probability of an event is given by an absolute square of a complex valued probability amplitude.
The amplitudes of two consecutive steps need to be multiplied.
\item[(b)] If the event can occur in different ways, and it is not possible to determine which of the alternatives is taken, 
the amplitudes for each way ought to be summed, and an interference pattern will emerge.
\item[(c)] If it is possible to determine, even in principle, whether one or the other alternative is taken, then the probabilities 
ought to be summed.
\item[(d)] The probabilities describe the state of the affairs when the experiment is \e{finished}.
\item[(e)] One never sums the amplitudes for different final states
 \cite{FOOT1}.
\item[(f)] Uncertainty Principle: one cannot determine which of the alternatives has been taken without 
destroying the interference pattern.
\cite{FOOT2}.
\end{itemize}

In Alice's case the event in question consists of preparing the system in a state $|a_k\ra$ at $t=t_0$, and then finding it in a
state $|c_n\ra$ at $t=t_2$. At time $t_1$, $t_0<t_1<t_2$, the system can be deemed to be in one of the states $|b_l\ra$, so there are,
in principle, $N$ different ways (paths), $\{c_n\gets b_l\gets a_k\}$, $l=1,...,N$, in which the event may occur. Notice that $c_n$,
$b_l$ and $a_k$ refer to states (lower case) and not to eigenvalues (upper case). By (a) the  corresponding amplitudes are
\begin{eqnarray}\label{1b}
\A(c_n\gets b_l \gets a_k) =\la c_n| \u^S(t_2,t_1)|b_l\ra\la b_l|\u^S(t_1,t_0)|a_k\ra,
\end{eqnarray}
and if no attempt is made to determine the state at $t=t_1$ the amplitudes should be added [cf. (b)], which yields
$\A(c_n\gets a_k)= \sum_{l=1}^{N}\A(c_n\gets b_l \gets a_k)=\la c_n|\u^S(t_2,t_1)|a_k\ra$.
\newline
If at $t=t_1$ Bob measures a $\B$ with degenerate eigenvalues, the amplitudes of the paths passing through the eigenstates of $\B$ corresponding to the same 
$B_m$ cannot be distinguished, and, by (b) and (c),  Eq.(\ref{1a}) follows. If some of the eigenvalues of $\C$ are also degenerate, the rule (e) prescribes adding the probabilities
Eq.(\ref{1a}) for all the eigenstates $|c_n\ra$ with the same eigenvalue $C_\mu$, which yields Eq.(\ref{2a}). 
\newline
According to \cite{FeynL}, the essence (the \e{only mystery}) of quantum mechanics is contained in the double-slit 
example. There a particle can reach a point on the screen via two holes, where the observed number of arrivals 
exhibits an interference pattern consisting of \e{bright} and \e{dark} fringes. Next we revisit the case in a slightly different context, and discuss it in some detail. 
\section{The double-slit case. The unspeakable and the  barely speakable}
To create a kind of a double-slit case in the context of quantum measurements, one needs three successive observations, similar to those
discussed in Sect. II. [Measuring just two operators, $\a$ and $\B$, won't do since, according to principle (e) in the previous section, there are no
interesting interference effects.] 
\newline
Allowing $\B$ to have only two different values, $B_1$ and $B_2$, and treating these alternatives as \e{slits}, 
creates two virtual paths connecting the \e{source} $|a_k\ra$ with $N$ \e{points on the screen}, represented by 
$|c_n\ra$. The corresponding amplitudes are  ($m=1,2$)
\begin{eqnarray}\label{1b}
\A(c_n\gets B_m \gets a_k) =\sum_{l=1}^{N} \Delta(B_m-\la b_l|\B|b_l\ra)\A(c_n\gets b_l \gets a_k). 
\end{eqnarray}
The \e{intensities} $W_n \equiv |\A(c_n\gets B_m \gets a_k)|^2$, corresponding to being in state $|a_k\ra$, measuring the value $B_m$, and finally
being in state $|c_n\ra$, may or may not show an \e{interference pattern}, depending on whether it is possible to determine which of the two
alternatives occurred. The simplest case is, of course, a two-level system, $N=2$, with only two \e{final locations}, $|c_1\ra$ and $|c_2\ra$
($\la c_2|c_1\ra=0$) [see Fig 2a].  For an initial state $|a_1\ra$,  special choice $|c_1\ra = \u^S(t_2,t_0)|a_1\ra$, makes $|c_1\ra$ and $|c_2\ra$, the bright and dark fringes,
respectively, since
\begin{figure}[h]
\includegraphics[angle=0,width=14cm, height= 8cm]{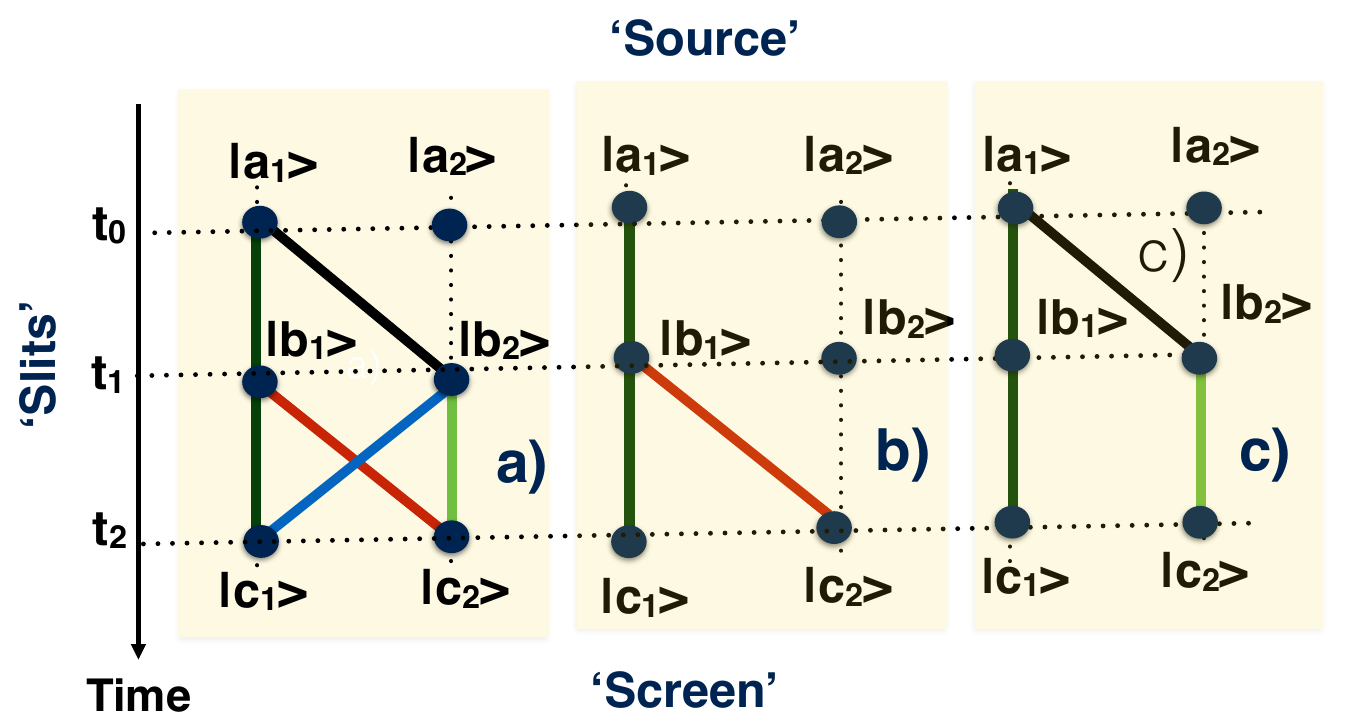}
\caption {a) The simplest double-slit problem. Intermediate, of $|b_{1,2}\ra$, and final, 
$|c_{1,2}\ra$, play the role of the \e{slits}, and \e{points on the screen}, respectively.
b) a special choice $|b_1\ra =\u^S(t_1,t_0)|a_1\ra$ eliminates two of the four virtual paths 
in a), and no interference is left in the system. c) similarly, choosing $|b_{1,2}\ra =[\u^S(t_2,t_1)]^{-1}|c_{1,2}\ra$
also reduces the number of paths to two.}
\label{Fig.4}
\end{figure}
\begin{eqnarray}\label{2b}
 W_1(\text{interference})=|\A(c_1\gets b_1 \gets a_1)+\A(c_1\gets b_2 \gets a_1)|^2=\la c_1|\u^S(t_2, t_1)|a_1\ra=1, \n
 W_2(\text{interference})=|\A(c_2\gets b_1 \gets a_1)+\A(c_2\gets b_2 \gets a_1)|^2=\la c_2|\u^S(t_2, t_1)|a_1\ra=0.
\end{eqnarray}
\subsection{Closing one slit}
In \cite{FeynL} Feynman pointed out the need for a \e{special language}, required to describe the double-slit conundrum. Firstly, the slit chosen
by the particle cannot be pre-determined by a \e{hidden} variable.
Were it so, closing one of the slits could only decrease the number of particles arriving at a given point (see also \cite{Bohr1}).
This, however, would be a wrong prediction since the closure makes the particles arrive at a previously dark spot.
Consequently, {\it it is not possible to say that the particle goes through one slit, and not the other}.
The same happens in our simple example. 
Although we cannot literally block one of the slits, it is possible to redirect the system, passing via $|b_1\ra$ or $|b_2\ra$
to different destinations, e.g., by  measuring $\B$, and counting separately the cases where the result is $B_1$ or $B_2$.
By (c) of Sect. III, knowing the value of $\B$ at $t_1$ makes the system arrive at a previously forbidden state $|c_2\ra$, 
\begin{eqnarray}\label{3b}
 W_1(\text{no\;interference})=|\A(c_1\gets b_1 \gets a_1)|^2+|\A(c_1\gets b_2 \gets a_1)|^2 < 1,\n
 W_2(\text{no\;interference})=|\A(c_2\gets b_1 \gets a_1)|^2+|\A(c_2\gets b_2 \gets a_1)|^2\ne 0,
\end{eqnarray}
which would not be possible if the patch chosen by the system were pre-determined at $t=t_0$.
In the general case, in order to \e{walk the logical tightrope} \cite{FeynL}, 
one cannot  say that the system 
found in $|c_n\ra$ at $t_2$ was {\it in one of the states $|b_m\ra$ at $t_1$}. [The only exceptions are the cases where the choice of $|b_m\ra$ leaves only one path connecting 
the initial and final states since the other amplitude vanishes, as shown in Figs. 2b and 2c.]
\subsection{Checking if the system passes through both \e{slits} at the same time}
Secondly, confirms \cite{FeynL}, it is not possible to say the a particle divides and  passes through both slits at the same time, 
since wherever one looks, one finds the entire particle, and not a fraction of it.  
To revisit the argument in the quantum measurements context, we assume, as before, that the initial and final states 
of our two-level system, $|a_1\ra$ and $|c_2\ra$, are accurately determined. For the probe $D^B$  we choose a von Neumann pointer \cite{vN}
with position $f$. If the probe, prepared in a normalised state $|G\ra$, briefly couples to the system at $t=t_1$ via
$\h_{int}= -i\B \delta(t-t_1)\;\partial_f$, the amplitude of finding a reading $f$ just after $t_2$ is given by
\begin{eqnarray}\label{1c}
\Psi(f)=G(f-B_1) \A(c_2\gets b_1 \gets a_1)+ G(f-B_2)\A(c_2\gets b_2 \gets a_1),
\end{eqnarray} 
where $G(f)\equiv \la f|G\ra$. 
Similarly, one can couple two probes to measure at $t=t_1$ the operators $\B=|b_1\ra \la b_1|$ and $\B'=|b_2\ra \la b_2|$. 
The amplitude of having readings $f$ and $f'$ becomes.
\begin{eqnarray}\label{1c}
\Psi(f,f')=G(f-1)G(f') \A(c_2\gets b_1 \gets a_1)+ G(f)G(f'-1)\A(c_2\gets b_2 \gets a_1).
\end{eqnarray}
Apparently, there are no scenarios in which both pointers are displaced, or both retain their original position. 
If the displacement of the pointer is to indicate the presence of the system in its respective path, one can only conclude that
the system cannot pass through both slits at the same time. An accurate measurement, where 
the width of $G(f)$ is much smaller than $1$ will always register the presence 
of the entire system in one of the pathways, apparently confirming this conclusion. 
But if the accuracy is such that $G(f)G(f-1)\ne 0$, the probability $|\Psi(f,f')|^2$ does contain 
an interference term $\R[G(f-1)G(f')G(f)G(f'-1)]\A(c_2\gets b_1 \gets a_1)\A(c_2\gets b_2 \gets a_1)]$, 
to which  both paths contribute simultaneously. 
One's inability to determine what happens to a quantum system in the presence of interference is reflected in the Uncertainty Principle (f) of Sect. III.
\newline 

\subsection{Fewer detections}
In \cite{FeynL} Feynman gives the following illustration of the importance of the Uncertainty Principle [(f) of Sect. III].
Suppose one determines the slit chosen in the double-slit experiment by sending a photon which is scattered only
if the particle passes through a given slit.
The destruction of the interference pattern on the screen is caused by the photon's collision with the particle. To minimise
the disturbance, one may try to lower  the intensity of light, but this will only reduce the number of photons, without diminishing
the strength of each  impact. As a result, all trials will fall into two categories: those in which the photon is scattered, and
the particle's path is known, and those where one cannot even say that the particle went through one slit and not the other
(cf. subsection A). The resulting pattern is a weighted sum of $W(\text{no\;interference})$ and $W(\text{interference})$. 
\newline
This too has a simple analogy in our discussion. The pointer's initial state $G(f)$ in Eq.(\ref{1c})  can be chosen to be a \e{rectangular window}
(see Fig. 3a),
\begin{eqnarray}\label{2c}
G(f)=1/\sqrt {\Delta f} \q \text{for}\; -\Delta f/2 \le f \le \Delta f/2, \q \text{and} \q 0\;\;\text{otherwise}.
\end{eqnarray}
\begin{figure}[h]
\includegraphics[angle=0,width=12cm, height= 8cm]{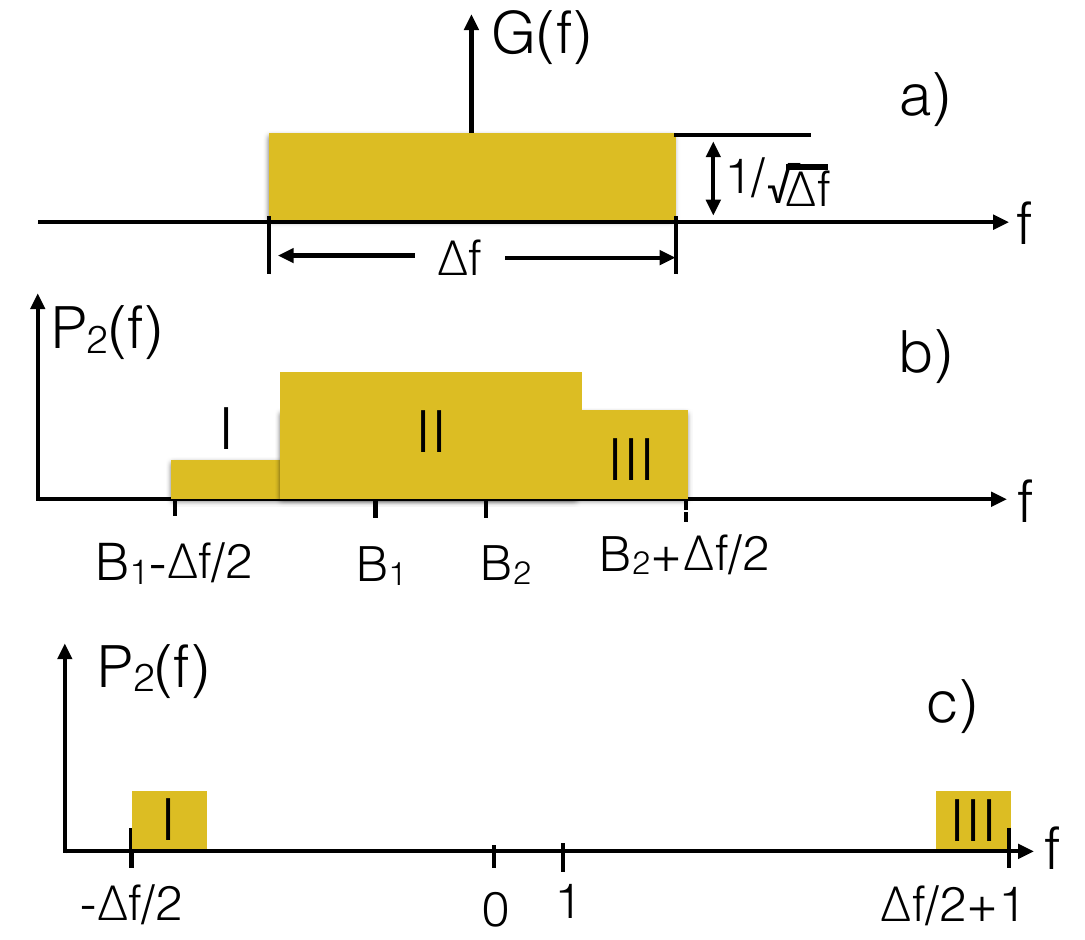}
\caption {a) The initial pointer's state $G(f)$ in Eq.(\ref{2c});
b) Probability distribution of the pointer's readings, $W(f)=|\Psi(f)|^2$, for a system found in $|c_2\ra$ at $t=t_2$
(schematic, arbitrary units). If a reading $f$ lies in regions $I$ or $II$, 
Bob knows that $\B$'s value was $B_1$ or $B_2$, respectively.
Nothing can be said about the value of $\B$ if the reading lies in the region II.
c) A projector $\B=|b_2\ra\la b_2|$ with the eigenvalues $0$ and $1$ is measured by a weakly coupled pointer,  $\Delta f >>1$,
in the case that the transition is forbidden,
$A(c_2\gets b_1 \gets a_1)=-\A(c_2\gets b_2 \gets a_1)$.
The readings are unusually large, $|f|>>1$, and Alice always knows the value of $\B$ at $t=t_2$. }
\label{Fig.1}
\end{figure}
The pointer becomes more  accurate as $\Delta f \to 0$, and $|G(f)|^2 \to \delta (f)$, in which case each trial yields 
either $f=B_1$ or $f=B_2$. For an inaccurate pointer, $\Delta f >|B_2-B_1|/2$, there is a region where the two $G$'s in (\ref{1c})
overlap (see Fig.3b). If Bob finds a reading $f$ inside that region, he can say nothing about the value of $\B$ (cf. subsection A).
Otherwise he knows the path chosen by the system.
As in the Feynman's example, the probability to find the system in a state $|c_n\ra$ at $t=t_2$ is a weighed sum
\begin{eqnarray}\label{3c}
W_n=\beta W_n(\text{no\;interference})+(1-\beta)W_n(\text{interference}),\q (n=1,2),
\end{eqnarray} 
where $\beta= 2|B_2-B_1|/\Delta f$, and $W_n(\text{interference})$ and $W_n(\text{no\;interference})$ are given by Eqs.(\ref{2b}) and (\ref{3b}), respectively.
Note that sending $\Delta f\to \infty$ is equivalent to introducing a new scaled variable  $f'= f/\Delta f$, 
which changed the coupling Hamiltonian to $\h_{int}= -i\B \delta(t-t_1)/\Delta f\;\partial_{f' }$.
Broadening $G(f)$ is, therefore, equivalent to reducing the coupling between the system and the pointer.
\newline
It is instructive to see what happens in the case of a previously forbidden transition (Fig. 3c), 
where $\A(c_2\gets b_1 \gets a_1)=-\A(c_2\gets b_2 \gets a_1)\equiv\A$, and Bob measures a projector $\B=|b_2\ra\la b_2|$.
For  $\Delta f > |B_2-B_1|$, $f$ may now lie in one of the intervals $[-\Delta f/2, -\Delta f/2+1]$ or $[\Delta f/2, \Delta f/2+1]$,
and in each case Bob knows that $\B$ had either a value $0$ (the system was in $|b_1\ra$), or $1$ (the system was in $|b_2\ra$).
The system reaches  $|c_2\ra$ with a probability $2|\A|^2/\Delta f <<1$, as the pointer has modified destructive interference
between the paths, allowing  it  to reach, on a few occasions,  the previously inaccessible final destination.
For $\Delta f >>1$ , Bob needs to wait a long time for an arrival in $|c_2\ra$, but in each 
such arrival the pointer's displacement is  very large, $\sim \pm \Delta f/2$. Note that this happens not because
the pointer has experienced an unusually large shift, but because it could be found not far from there even before the measurement
was made. 
\subsection{Fuzzy detections}
Thus, continues Feynman \cite{FeynL}, the way to observe the particles without disturbing them considerably
may be to increase the photon's wavelength, rather than decrease their number. A red coloured photon carries a smaller momentum 
than a blue one, so the jolt produced on the particle should also be smaller. However, if the wavelength $\lm$ were to exceed the distance
between the slits, $d$, one would only see a \e{fuzzy} flash, unable to decide which of the two slits it came from.  
This is another illustration of the Uncertainty Principle (f): \e{an apparatus, capable of determining which way a particle goes,
cannot be so delicate so as not to disturb the interference pattern in an essential way  \cite{FeynL}.}
\newline
Such a fuzzy detection can be mimicked by using a pointer prepared in a Gaussian (rather that in a rectangular) state,
\begin{eqnarray}\label{1d}
G(f, \Delta f)= \left(\pi\Delta f^2/2\right)^{-1/4} e^{-f^2/\Delta f ^2},
\end{eqnarray}
where $\Delta f $ exceeds the difference of the measured values, and the condition $\lm > d$ is replaced by $\Delta f > |B_2-B_1|$.
\newline
Consider again the situation where one measures an operators $\B=|b_1\ra \la b_1|$. The
probability of finding a reading $f$, conditional on post-selection in $|c_2\ra$, 
is given by
\begin{eqnarray}\label{2d}
W(f)= |G(f-1)\A(c_2\gets b_1 \gets a_1)|^2+|G(f)|\A(c_2\gets b_2 \gets a_1)|^2+\n
2\R[G(f-1)G(f)\A^*(c_2\gets b_1 \gets a_1)\A(c_2\gets b_1 \gets a_1)].\q\q
\end{eqnarray}
Even though there are two scenarios the system may follow, the last interference term clearly prevents one from knowing
which of them has actually been realised if the measurement is fuzzy. 
The term  disappears if the measurement is sharp $\Delta f << |B_2-B_1|$, and the system is always seen to choose one of the two paths. 
\newline
One might object by pointing out that we kept changing the subject. By sending $\Delta f \to 0$ Bob perturbs the system considerably,
while one only wished to know what happened to the {\it unperturbed} system. There is an orthodox answer to that too. The wish cannot be
granted, or the Uncertainty Principle (f), necessary for the very existence of the quantum theory \cite{FeynL}, would be violated. 
It appears that the \e{which way?} question simply cannot be answered in the presence of interference. 
\newline
We could finish the discussion with the adaptation of the general rules of \cite{FeynL} to the needs of the measurement theory, were it not for one additional circumstance.}
{
\section{The \e{Weak measurements}}
In an attempt  to study the properties of an (almost) unperturbed system, Alice may send $\Delta f $ in Eq.(\ref{1d}) to infinity.
The first thing she would notice is that if the final position of the pointer is representative of the value of the measured $\B$, this value is,
indeed, indeterminate, since $f$ can lie almost anywhere. The answer to a question which should have no answer can be \e{anything at all}. She can
also expect that no other useful information can be extracted from the very broad distribution of the pointer's reading. It is only reasonably  to expect
that the final {\it mean} reading of an infinitely inaccurate pointer would also be zero, just as was the initial one. 
\newline
This is, however, not the case.
In \cite{SPIN100} the authors have shown (using a different language and notations) that the average pointer's reading in the limit $\Delta f \to \infty$
is given by 
\begin{eqnarray}\label{1e}
\la f\ra \equiv \frac{\int f |\Psi(f)|^2 df}{\int |\Psi(f)|^2 df}{\xrightarrow[\Delta f \to \infty ] {}}\R\left [\sum_lB_l\alpha_l
 \right ] \equiv \R[\la \B\ra_W],
\end{eqnarray} 
where 
 \begin{eqnarray}\label{2e}
\alpha_l =\frac{\A(c_2\gets b_l \gets a_1) }{\sum_{l'} \A(c_2\gets b_l' \gets a_1) }
\end{eqnarray}
is the path amplitude, renormalised to a unit sum over the paths leading to the same destination $|c_2\ra$, $\sum_l \alpha_l=1$.
 The complex valued quantity $\la \B\ra_W$ was called \e{the weak value (WV) of $\B$}.
Its imaginary part, $\I [\la \B\ra_W]$, can be obtained if the pointer's mean final momentum is measured instead \cite{SPIN100}, (see also \cite{DSpla1}). 
{\r In practice this means that after repeating his measurements $K>>1$ times, Bob will be able to deduce the value in the r.h.s. of Eq.(\ref{1e})  
 sufficiently accurately, even if the uncertainty of each trial, $\Delta f$, is as large as he wants, and the interference between the system's paths is affected as little 
 as he wants. [Bob's task, although possible in principle, will rapidly become impractical. Indeed, by the Central Limit Theorem \cite{CLT}, 
the number of trials (copies of the system)  will have to grow as $K\sim \Delta f^2/\R[\la \B\ra_W]^2$. ]}
 \newline
This result should be of interest to anyone wishing  to challenge the conclusions of  Refs. \cite{FeynL}, \cite{FeynC}, and \cite{FeynH}.
Contrary to one's expectations, a concrete information clearly related to the intermediate condition of an (almost) 
unperturbed system can be obtained experimentally. 
Does measuring $\la f \ra$ in this manner allow Bob to speak about the things Feynman thought unspeakable? 
\subsection{\e{Weak values} and the Uncertainty Principle}
Apparently not. The Uncertainty Principle of \cite{FeynL} does not forbid knowing the values of the probability amplitudes.
It does, however, forbid to make assumptions about the system's past where the scenarios one wishes to distinguish continue to interfere. 
Nothing in the Principle stops Bob from learning  the values of either the amplitudes (\ref{2e}), or of their various combinations (\ref{1e}). 
The question is, what conclusions is he allowed to draw from the \e{weak values} once they have been measured? 
\newline
Before addressing it,  it may be useful to demonstrate that 
there always exists 
a transition in which the weak value of any chosen operator $\B$ equals any desired complex number $Z$. Indeed, there are infinitely many choices
of $\alpha_l=\A_l/\sum_{l'}\A_{l'}$, such that $\la \B\ra_W=\sum_\l B_l \alpha_l$ equals $Z$. If a particular set $\alpha_i(Z)$ has been chosen, 
the path amplitudes $\A_l$ must satisfy
 \begin{eqnarray}\label{1i}
\sum_{l'=1}^N [S_{ll'} - \delta_{ll'}]\A_{l'}=0, \q (S_{ll'}\equiv \alpha_l\q \text{for all}\q l').
\end{eqnarray}
Since $\sum_{l=1}^N\alpha_l =1$, the matrix in the square brackets is degenerate, and the system (\ref{1i}) has a particular solution
$\A_l=\alpha_l/\alpha_1$. 
\newline 
If only three quantities are measured, the amplitudes are given by 
 \begin{eqnarray}\label{2i}
\A_l=\la c_n|\u^S(t_2,t_1)|b_l\ra\la b_l|\u^S(t_1,t_0)|a_k\ra\equiv \la b_l|c_n(t_1)\ra^* \la b_l|a_k(t_1)\ra,  
\end{eqnarray}
and for any $|a_k\ra$
one finds a (yet unnormalised) state $|c_n(t_1)\ra$, $\la b_l|c_n(t_1)\ra=\alpha_l^*/[\alpha_1\la b_l|a_k(t_1)\ra]^*$. Normalising $|c_n\ra$ to unity yields a final state $|c_n\ra$ such that, for a given initial state $|a_k\ra$, $\la \B\ra_W$ equals the desired value $Z$, 
 \begin{eqnarray}\label{3i}
 |c_n\ra=\u^S(t_2,t_1)|c_n(t_1)\ra, \q 
\la b_l|c_n(t_1)\ra = \frac{\alpha_l^*(Z)}{\la b_l|\u^S(t_1,t_0)|a_k\ra}\left[\sum_{l'=1}^N \frac{|\alpha_{l'}(Z)|^2}{| \la b_{l'}|\u^S(t_1,t_0)|a_k\ra|^2} \right ]^{-1/2}.\q
\end{eqnarray}
The indeterminacy implied by the Uncertainty Principle is restored, albeit at a higher level,
where all possible transitions are considered. The weak value of a $z$-component of a spin-$1/2$ 
can  be found to have a value $\la \hat \sigma_z \ra_W=1/2$, as well as 
$\la \hat \sigma_z \ra_W =1.2345*10^{19} - 6.789*10^{28} i$. Both results are equally meaningful or, if one prefers, equally meaningless, as we discuss next. 
\subsection{Interpretation of the \e{weak values}}
This is the most contentious part of our discussion.
Enter Carol who, unlike Alice, 
suspects that Feynman's narrative \cite{FeynL} it too restrictive, and there is more to be learnt through the \e{weak measurements} described above.
\newline
{\bf C}: You forbid saying anything about the value of $\B$ half way through the transition. I put it to you that this weak (WV) value is given by Eq.(\ref{1e}). The authors of \cite{SPIN100} have shown that 
Bob can measure a component of a spin-$1/2$ and give you a number $100$. Here is the answer to a question which  you thought could have no answer. 
\newline 
{\bf A}: By the same logic, the WV of a projector $\B=|b_1\ra\la b_1|$ should be the probability of the system taking  that path.
I can always find a transition [cf. Eq.(\ref{3i})] where $\la \B \ra_W=-5i$. This is not a probability. A probability is related
to the number of occurrences, and always has value between $0$ and $1$. I do, however, have a different name for this result:
it is a probability amplitude (\ref{2e}).
\newline 
{\bf C}: All right, one cannot substitute the relative amplitudes for probabilities, but maybe they be interpreted as \e{occupation numbers}, as was
suggested in \cite{OCCUP}. Let us say that the \e{weak value}  of a projector $\B=|b_l\ra\la b_l|$, $\la \B \ra_W=\alpha_l$, represents,
in some sense,  the \e{number of particles} in the $l$-th path (the $l$-th arm of an interferometer, if an optical realisation of the experiment is
considered). These weak values can be measured simultaneously, obey a \e{self-consistent logic}, and should lead to a \e{deeper understanding of the nature of quantum mechanics} according to \cite{OCCUP}.
\newline 
{\bf A}: I am afraid, you would have the same problem: $-5i$ is by no stretch of imagination a valid number of particles 
(or particle pairs as proposed in \cite{OCCUP}). It would be even worse for a \e{dark} final state, where $\sum_{l} \A(c_2\gets b_l \gets a_1)=0$,
unless you want 
to tell 
Bob that infinitely many particles in each pathway conspire so that 
nothing arrives at the final destination $|c_2\ra$. Besides, a WV of a projector is not a good local indicator, and should not be used to quantify 
what happens in a particular arm of the interferometer (see Appendix A). The WV of $\B=|b_l\ra\la b_l|$ depends on the amplitudes of all pathways leading to the same final state [cf. Eq.(\ref{B5})], in a way  its accurate  \e{strong} value does not [cf. Eq.(\ref{B3})].
And, by the way, the \e{self-consistent logic} \cite{OCCUP} enjoyed by the weak values is but the property that
the relative amplitudes (\ref{2e}) always add up to unity. 
\newline   
{\bf C}: Let us forget about the numerical values of the WVs, be they real or imaginary, and simply use them as indicators of the particle's presence in particular part of the system, as was suggested in \cite{PHOT2}. One couples  a weak pointer to a particular system's state (or a part of the interferometer).   
Surely, if the pointer does not move (the weak value of the corresponding projector is $0$), the system never visits that part of the Hilbert space?
If, on the other hand, it did move, the system must have been there. What can possibly go wrong with relying on such a \e{weak trace}?
\newline
{\bf A}: This is the same as saying that the system does not use a pathway whose amplitude is zero.
But the pathways  can be combined, and you  will be in danger of finding the particle in parts while no finding it in the whole. 
This is not too different from Feynman's dilemma of Sect.3A, whereby one finds more particles arriving at the screen 
if one slit is closed. We will discuss this case next. }
{
\section{The three-slit case.}
With three rather than just two dimensions it comes a possibility of measuring operators with degenerate eigenvalues, which we will explore next. There are now  three paths, connecting $|a_1\ra$ with $|c_1\ra$, which, to
shorten the notations, we will denote $\{l\}\equiv \{c_1 \gets b_l\gets a_1\}$, and three amplitudes, $ \A_l\equiv\A(c_1\gets b_l\gets a_1)$, $l=1,2,3$ (see Fig.4a).
We will consider two commuting projectors,
\begin{eqnarray}\label{2g}
\B=|b_1\ra\la b_1|,\q\q\q\q\q B_1=1, \q B_2=B_3=0,\q\q\q\q\n
\B'=|b_1\ra\la b_1|+|b_2\ra\la b_2|, \q B'_1=B'_2 =1, \q B'_3=0,\q\q\q\q
\end{eqnarray}
designed to question the presence of the system, at $t=t_1$, in the state $|b_1\ra$, and in the sub-space spanned by  $|b_1\ra$ and $|b_2\ra$, 
respectively. Following  \cite{3BOX}, we tune the system to ensure 
\begin{eqnarray}\label{3g}
\A_1=-\A_2=\A_3\equiv A,\q\q (\alpha_1 =-\alpha_2=\alpha_3=1).
\end{eqnarray}
If no measurements are made, the probability of the system arriving at $|c_1\ra$ at $t=t_2$ is $W_1=\left |\sum_{l=1}^3 A_l\right |^2= |A|^2$.
\subsection{No paradox so far}
An accurate measurement of $\B$ always finds the system in the state $|b_1\ra$ (we omit  the dependence on $|a_1\ra$ and $|c_1\ra$)
{\b 
\begin{eqnarray}\label{4g}
P(B=1)
=|\A_1|^2/(|\A_1|^2+|\A_2+\A_3|^2) =1.
\end{eqnarray}}
However, an accurate measurement of $\B'$ never finds it in the sub-space containing $|b_1\ra$, 
{\b
\begin{eqnarray}\label{5g}
P(B'=1)
 =|\A_1+\A_2|^2/(|\A_1+\A_2|^2+|\A_3|^2) =0.
\end{eqnarray}}
In both cases the probability of a successful post-selection in $|c_1\ra$ is, as before, $W_1= |A|^2$.
Although the authors of \cite{3BOX} found this \e{paradoxical}, there is nothing  particularly surprising.
In Sect. II we went to some length stressing the need for employing a probe which would carry the record 
of the outcome obtained at $t=t_1$ until the experiment is finished just after $t_2$.
Measuring $\B$ and $\B'$ requires different probes ($-i\partial_f \B\ne -i\partial_f \B'$) which affect the system,
expected later to reach the same final state, in different ways.
The resulting statistical ensembles are, therefore, {\it different}, and there is no conflict between  Eqs.(\ref{4g}) and (\ref{5g})
\cite{DSpath}. 
\newline
Measuring $\B$ and $\B'$ one after another also shows nothing unusual. With two intermediate measurements there are 
$9$ paths, of which only three have non-zero amplitude (we put $\h^S =0$)
\begin{eqnarray}\label{6g}
\A(c_1\gets b_{l'} \gets b_l\gets a_1)=\la c_1|b_{l'}\ra\la b_{l'}|b_{l}\ra\la b_{l}|a_1\ra= A\delta_{ll'}.
\end{eqnarray}
There are also four possible outcomes, $B=0,1$, $B'=0,1$. The outcome $(B=1,B'=0)$ would indeed be problematic, 
indicating the presence of the system in the path $1$ but not in the union of the paths $1$ and $2$. 
However, this scenario is never realised, as the corresponding probability vanishes, $P(B=1,B'=0)=0$ (see Fig.4b). 
Each of the three remaining outcomes allow to identify the paths taken by the system, and by (c) of Sect. III, 
one has 
{
\begin{eqnarray}\label{7g}
P(0,0)=P(0,1)=P(1,1) = |A|^2/(|A|^2+|-A|^2+|A|^2)=1/3. 
\end{eqnarray}}
We note also that, just like in Feynman's two-slit example of Sect.4A, determination of  the path taken by the system increases the
odds on its arriving in the state $|c_1\ra$, $W_1=\sum_{l=1}^3 \left |A_l\right |^2= 3|A|^2$. One cannot, therefore, say
that an unobserved system is in one of the states $|b_l\ra$, and not in the other two.  
By the same token,   whenever some of the paths are not resolved by a measurement (the paths $\{1\}$ and $\{2\}$ in our 
example), it should be impossible to say that the system chooses one of them and not the other [cf. Eqs.(\ref{4g}) and (\ref{5g})].
Next, as promised,  we will bring  the \e{weak values} into the discussion.
\begin{figure}[h]
\includegraphics[angle=0,width=13cm, height= 12cm]{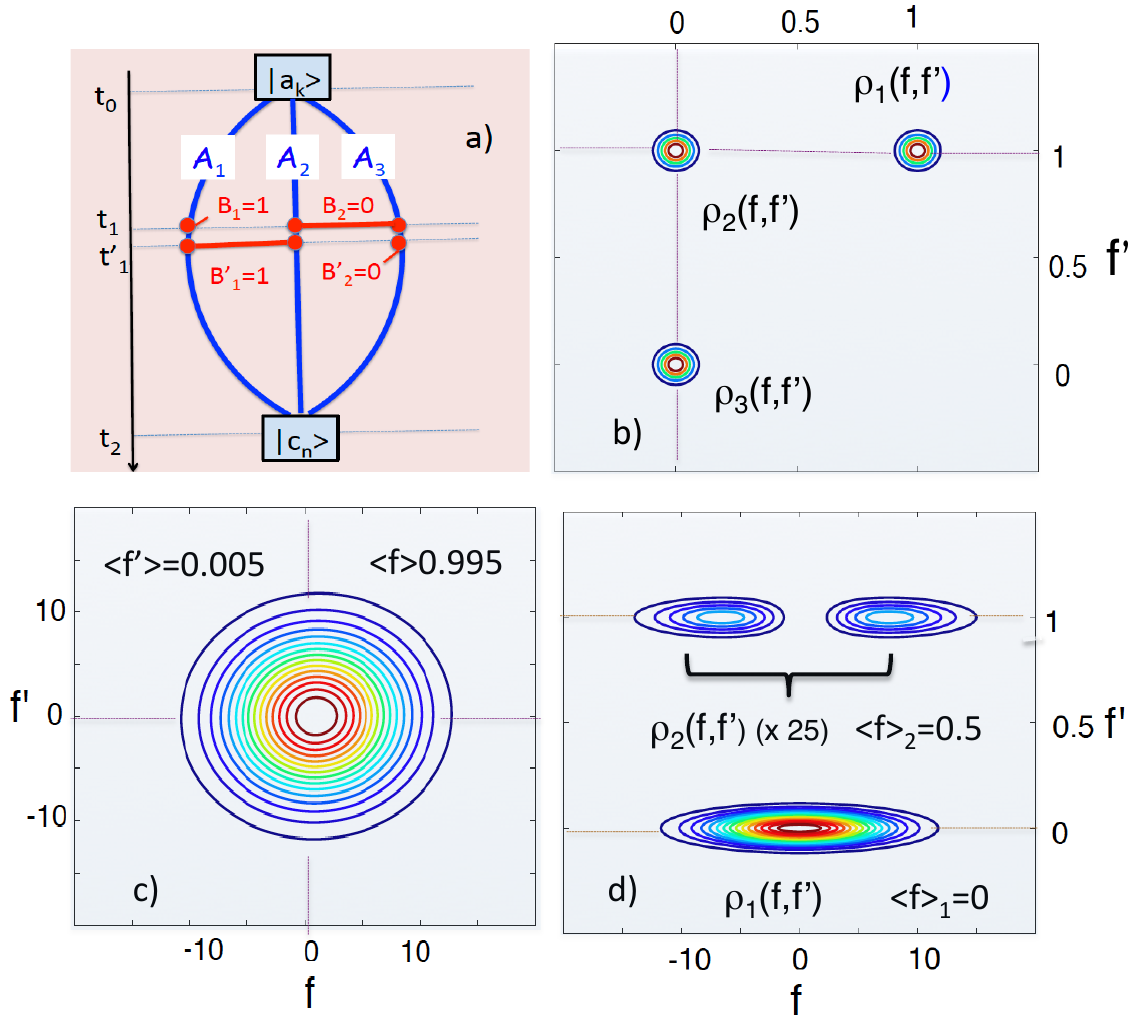}
\caption {a) Three virtual pathways with the amplitudes
$\A_1=-\A_2=\A_3$. At $t=t_1$ Bob measures the operator $\B$ in Eq.(\ref{2g}), 
and at $t=t_1'$, immediately after,
the operator $\B'$. 
What happens depends on the accuracy of Bob's pointers.
b) The distribution of the pointer readings, $\rho(f,f')$ when both pointers are accurate (strong), $\Delta f= \Delta f'=0.1$.
c) The same distribution when both pointers are inaccurate (weak), $\Delta f= \Delta f'=10$. 
d). The first pointer is accurate, $\Delta f =0.1$, and the second one is weak, $\Delta f' =10$.  }
\label{Fig.4}
\end{figure}
\subsection{An unlikely \e{cat} }
To add panache to the story, we will refer to the system as a \e{cat}, 
promote the paths $\{1\}$ and $\{2\}$ to a \e{one-bedroom  flat}, and call the remaining pathway $\{3\}$
the \e{yard}. Our aim, we recall, is to divine the behaviour of an unobserved and, therefore, unperturbed cat.
To do so we will relay on the existence or absence of the \e{weak traces}, non-zero \e{weak values} of the corresponding projectors. 
Using Eq.(\ref{3g}) we find that the weak pointer looking for the cat in the bedroom $\{1\}$ has indeed \e{moved}, 
 \begin{eqnarray}\label{8g}
 \la \B \ra_W = B_1\alpha_1+B_2(\alpha_2+\alpha_3)=1\ne 0.
\end{eqnarray}
The weak pointer looking for it in the entire flat, however, did not (see Fig.4c)
 \begin{eqnarray}\label{9g}
\la \B' \ra_W = B'_1(\alpha_1+\alpha_2)+B_3\alpha_3=0, 
\end{eqnarray}
and we have a \e{paradox}: {\it A quantum cat can be in a part of the flat, and yet not in the flat as a whole.} 
\newline
This  puzzle is, however, not entirely our own.
A similar \e{paradox} was described in \cite{PHOT2}, where three successive WMs yielding $\la \B' \ra_W=0$,  $\la \B \ra_W\ne 0$ and again  $\la \B' \ra_W=0$ gave the impression that a quantum system can suddenly appear in a the state $|b_1\ra$, without being there immediately before and immediately after. An optical realisation of the experiment led the authors of \cite{PHOT2} to conclude that \e{the past of the photons is not represented by continuous trajectories}. 
Next we ask whether a similar interpretation of the results (\ref{8g}) and (\ref{9g}) can lead Bob to a wrong prediction. 
\section{The three-slit  case. How to make a wrong prediction}
Bob the practitioner is duly surprised, and wants to make further predictions based on the cat's strange behaviour. 
He may be thinking: the weak pointer measuring $\B'$ \e{has not moved}, $\la \B' \ra_W=0$, because it has not interacted with the cat, who is absent 
from the flat. Surely, if the cat is really absent, a closer look inside the flat would not affect it at all. Thus, a more accurate measurement of 
$\B'$  should not change anything for the weak pointer measuring $\B$.
Bob improves the accuracy of the second measurement by sending $\Delta f' \to 0$, leaves the measurement 
of $\B$ \e{weak}, and expects still to find the trace of the cat in the bedroom $\{1\}$. 
\newline 
This is, however, not what he finds. Bob's accurate measurements of $\B'$ have created two 
\e{real} (as opposed to \e{virtual}) scenarios for the system. It can reach the final state 
$|c_1\ra$ via the path $\{3\}$, with a probability {\b $|\A_3|^2$}.
Alternatively it can reach it via a scenario in which 
the paths $\{1\}$ and $\{2\}$, not resolved by the first Bob's measurement of $\B$, continue to interfere. The second possibility is a double-slit problem in its own right, and a one in which 
virtual 
paths  lead to a \e{dark fringe}, since $\A_1=-\A_2$. The probability of taking this route is, 
therefore, zero, {\b $|\A_1+\A_2|^2=0$}. The mean reading of the weak pointer measuring $\B$, conditional on finding the system in $|c_1\ra$, is
{\b  
 \begin{eqnarray}\label{1h}
\la f \ra =\frac{\R\left [\frac {B_1\A_1+B_2\A_2}{\A_1+\A_2}\right ]|\A_1+\A_2|^2 +B_3|\A_3|^2 }{ |\A_1+\A_2|^2+|\A_3|^2}=0, 
\end{eqnarray}}
so there is no cat in the flat, no longer a \e{weak trace} of the cat in the bedroom and, therefore, no paradox.
\newline
To see how the result (\ref{1h}) is obtained, consider  a measurement of $\B'$ to a good, yet finite accuracy, $0<\Delta f'<<1$
Bob's results are shown in Fig. 4d.
The probability distribution of the readings 
consists of  two disjoint parts, $\rho(f,f')=[\om_1+\om_2]/\int(\om_1+\om_2)df df'\equiv \rho_1+\rho_2$ (see Appendix B) 
where
 \begin{eqnarray}\label{1ha}
\om_1(f,f') =|G(f,\Delta f)|^2|G(f',\Delta f')|^2|\A|^2,\q\q\q\q\q\q\q\q\q \q \n
\om_2(f,f') =|G(f-1,\Delta f')-G(f,\Delta f')|^2 |G(f'-1,\Delta f')|^2|\A|^2|. 
\end{eqnarray}
The first part corresponds to the cat not found in the flat, and shows no  trace of the cat in the bedroom, $\la f\ra_1=
\int f \rho_1(f,f')dfdf'=0$.  The second part finds the cat in the flat, and there is also a trace of it in the bedroom (the meters are Gaussian),
  $\la f\ra_2=\int f \rho_1(f,f')dfdf'{\xrightarrow[\Delta f \to \infty ] {}}1/2$. However, the probability of the latter outcome,
  which occurs only  because the second Bob's pointer slightly upsets otherwise perfect cancellation between the paths $\{1\}$ and $\{2\}$,
  vanishes as $\Delta f \to \infty$
  $\int \rho_2(f,f')df df' \sim 1/\Delta f^2{\xrightarrow[\Delta f \to \infty ] {}}0$ (see Appendix B), and Bob only gets the outcomes where neither of the two pointers \e{has moved}. 
\newline
Alice and Carol again:

{\bf C}: A quantum system can be absent from a subspace of its Hilbert space, yet simultaneously present in a smaller 
part of that subspace. In an optical experiment a photon can be present in one arm of the interferometer, but absent 
from a larger part of the setup, containing  the arm in question. 
\newline
{\bf A}: And how do you know that?
\newline 
{\bf C}: A weak pointer moves (albeit on average)  if I look in a smaller subspace, but not if I look in the larger one. 
\newline 
{\bf A}: Your \e{paradoxical} conclusion is predicated on the link between the movement of the pointer and the actual presence 
of the system at the specified location. But you can also look at your result differently. You are simply saying that the system is simultaneously 
present in all the pathways were the corresponding amplitudes, {\b $\A_l$},  does not vanish. If so, in a double-slit experiment 
a particle passes via both slits at the same time. In the orthodox view \cite{FeynL}, \cite{FeynC}, \cite{FeynH} you can always say that an 
${\b \A_l}\ne 0$.
You cannot, however, conclude from this that the system, or a part of it,  {\it is} there. 
\newline 
{\bf C}: And why is that? 
\newline 
{\bf A}:
Because a more accurate measurement may find the entire system elsewhere, or destroy the \e{paradox} as was discussed in above example. Your prediction will be proven wrong.
\newline 
{\bf C}: But you said yourself that an accurate measurement will disturb the system, and I am taking about the experimental 
evidence of  an unperturbed system's behaviour. Perhaps the weak measurements, unknown to  the authors of \cite{FeynL}, \cite{FeynC}, \cite{FeynH} at the time, conveniently filled the gap left in their reasoning? 
\newline 
{\bf A}:  The authors of \cite{FeynL}, \cite{FeynC}, \cite{FeynH} have an answer to this. The fact that one can deduce the values of  the amplitudes experimentally is of no fundamental importance.
The amplitudes are always known to a theorist, and the orthodox contention is that they cannot be used beyond what is
covered by the rules of Sect. III.  In  \cite{FeynC}, Feynman admits that one can {\it say} something about what happens to the system in the presence of interference, but  needs \e{to stop thinking immediately and make no deductions from it}. 
You can {\it say} that there are particles passing through each slit yet never arriving at a dark fringe on the screen.
You  can {\it say}  that photons have discontinuous trajectories \cite{PHOT2}, or that a cat both is and is not inside the flat, 
but that is all. \e{Physicists prefer not to say it , rather than stop thinking at the moment.} \cite{FeynC}.

\section{The four-slit case. \e{Quantum Cheshire cats} }
Adding the fourth dimension allows one to consider a pair of two-level systems.
Following \cite{CAT1} call the first one a \e{Cheshire cat} (sic!), and the second one the cat's \e{grin}.
The cat can be either \e{on the right}, or \e{on the left}, i.e.,  in states $|L\ra$ or $|R\ra$, respectively.
The grin can be \e{up} or \e{down}.
The four possible states of the joint system, therefore, are
 \begin{eqnarray}\label{1j}
|b_1\ra=|L\ra|\+\ra,\q |b_2\ra=|L\ra|\-\ra,\q |b_3\ra=|R\ra|\+\ra,\q |b_4\ra=|R\ra|\-\ra.
\end{eqnarray}
There are also two operators, diagonal in the basis (\ref{1j}),
\newline
$\B=|L\ra\la L|$, with the eigenvalues $(1,1,0,0)$, and
\newline
$\B'= |R\ra\la R|\otimes(|\+\ra\la\+| - |\-\ra\la \-|)$, with the eigenvalues $(0,0,1,-1)$, 
\newline
where the operator $|\+\ra\la \+| - |\-\ra\la \-|$, akin to the $z$-component of a spin-$1/2$, $\hat \sigma_z$, represents the \e{cat's} \e{grin}.
One can always choose (see Sect. VA) the four amplitudes $ \A_l \equiv \A(c_n \gets b_l\gets a_k)$, $ l=1,2,3,4$ as
 \begin{eqnarray}\label{3j}
\A_1 =-\A_2=\A_3=\A_4, \q (\alpha_1 =-\alpha_2=\alpha_3=\alpha_4 =1/2).
\end{eqnarray}
We leave it to the reader to check what happens if the above operators are measured accurately, together or separately, 
and note only that the \e{weak values} of both operators vanish,
 \begin{eqnarray}\label{4j}
\la \B \ra_W =0,\q
 \la \B' \ra_W =0.
\end{eqnarray}
\newline
To follow the story one has 
(a) to agree that a null  \e{weak value} of $\B$ 
 means that the cat is {\it not} on the left ,
  and
(b) to accept that  a null \e{weak value} of $\B'$ 
means that the cat's grin is {\it not} on the right.
There is also a tacit assumption that if something is not in one place, it must be in the other, 
since neither
the cat nor the grin are expected to disappear. 
Thus, one may conclude
  that {\it the cat (on the right) has been \e{separated} from its grin (on the left),}
and consider other possible \e{separations}  mentioned in the second of our epigraphs.
Further discussion of this \e{paradox} can be found in \cite{CAT1},\cite{CAT2}.
\newline
Our purpose is to question the principle.
In Eq.(\ref{4j}) $\la \B \ra_W$ vanishes because  $B_1=B_2$ and $\A_1 =-\A_2$. This situation was considered in some detail 
in the previous two Sections. The case of $ \la \B' \ra_W =0$ is slightly different. Now the path amplitudes 
are the same,  $\A_3 =\A_2$, but the two eigenvalues have opposite signs, $B'_3=-B'_4$.
\newline
To clarify the situation,
Bob (aware of Carol's objections to 
using strong measurements) may decide to add yet another \e{weak} pointer, measuring $\B''=|L\ra\la L|\otimes|\+\ra\la\+|$. 
Whoever agrees with (a) and (b) must also accept that $\la \B'' \ra_W$ indicates the presence of an \e{upwards  grinning} cat  at the left location. Having been told that there is no cat on the left, Bob expects 
to obtain $\la \B'' \ra_W=0$, but finds instead $\la \B' \ra_W=1/2$.
Similarly, weakly measuring $\B'''=|R\ra\la R|\otimes|\+\ra\la\+|$, he finds the \e{up} component of the grin on the right, since
$\la \B''' \ra_W$ is also $1/2$. 
\newline
The story \e{told by the simultaneous weak values}  is now as follows:

(i) there is no cat on the left hand side of the setup,

(ii) yet there is a cat there,  provided its grin is \e{up},

(iii) there is no grin on the right hand side, 

(iv) yet there is a grin, provided it is \e{up}. 
\newline
Alice and Carol again. 
\newline 
{\bf A}: You told Bob that there is no grin on the right  hand side, but he finds a part of it there. Another wrong prediction?
\newline 
{\bf C}:
Au contraire, you just discovered another quantum mechanical paradox: Bob can have  the cat in one place, 
the cat's grin elsewhere, and still a bit of the grin where the grin is not. Have you two thought of a publication?
\newline 
{\bf A}:  All we know is that  Bob's \e{weak measurements} have confirmed the relations (\ref{3j})
between the four amplitudes in question. But there is no surprise, Bob has set up his experiment ensuring that 
$\alpha_1 =-\alpha_2=\alpha_3=\alpha_4$. 
Your \e{paradox}
disappears the moment you stop interpreting Eqs.(\ref{3j}) the way you do.
\newline 
{\bf C}: And how should I interpret them?
\newline 
{\bf A}: You should not interpret them at all, lest you create spurious \e{paradoxes}.
\newline 
{\bf C}: What if I like paradoxes?
\newline
The next Section contains our conclusions. }
{
\section{Conclusions and discussion}
In summary, our aim was compare two apparently contradictory approaches to the description of quantum  behaviour.
The first one, which we called the orthodox, can be applied to quantum measurements as follows.
An experimentalist wants to measure three (or more) quantities one after another. 
He/she needs probes (pointers) to read off the past measured values at the end of experiment. 
A theorist  describes the system in terms of virtual paths connecting the states in the 
system's Hilbert space $\mathcal H_S$, and the corresponding probability amplitudes $\A_l$. If the equipment
allows to determine the path taken by the system in each trial,  absolute squares 
of the amplitudes, $|\A_l|^2$, predict the relative frequency (probability) with which a particular path will occur 
after many  trials.  A quantity $\bf B$ is represented by an operator $\B$ acting in $\mathcal H_S$.
Its mean value, $\la B\ra$, is then obtained by multiplying each path probability by the corresponding eigenvalue of the operator $\B$, adding up the results, and normalising  the sum as appropriate.
\newline
If the paths taken cannot be told apart, the amplitudes must be added, and one can neither say that a particular scenario was  realised, nor that all the scenarios were realised at the same time. According to \cite{FeynL}, \cite{FeynC}, \cite{FeynH}, either assumption about the system's past 
 would lead to wrong predictions, or contradictions. 
In particular, not  even the mean value of $\bf B$ can be determined at a time between two observations. 
Moreover, there is no deeper explanation of the law of adding amplitudes,
which must be accepted with all restrictions  it imposes upon what can be known. These restrictions amounts to the Uncertainty Principle, a centrepiece  of quantum description, without which \e{quantum mechanics would collapse} \cite{FeynL}.
\newline
The Principle  appears to leave an unpleasant  void at the heart of the theory, and a desire to fill it can only be natural. The task is, however, difficult 
and, if attempted, must end either in an important new development, or in a misunderstanding of similar proportions. 
Feynman, for his part,  was in no doubt that it would end in a \e{blind alley, from nobody has yet escaped} \cite{FeynC}.
\newline
A different approach aims at describing a quantum system between measurements \cite{COMPL} and, can, therefore,
 be seen as an attempt to fill the gap left by the Uncertainty Principle. 
For the purpose of our discussion it suffices to note that the method relies on experimentally accessible \e{weak value} (WV),  $\la B \ra_W$, a sum of the probability amplitudes, weighted by the eigenvalues of the measured
operator $\B$, 
 \begin{eqnarray}\label{1x}
 \la B \ra_W = \sum_l B_l \alpha_l, \q
\alpha_l =\frac{\A_l }{\sum_{l'} \A_{l'} }.
\end{eqnarray}
(more technical details are given in Appendix A).
The amplitudes $\alpha_l$ are renormalised to a unit sum, and there always exists  a transition, in which $\la B \ra_W$ takes any desired complex value (cf. Sect. V). As yet, there is no contradiction with the Uncertainty Principle which allows knowing probability amplitudes
regardless of whether or not the system is being observed. It remains, however,  to clarify the meaning of the result (\ref{1x}). Different propositions have been put forward by the authors of \cite{SPIN100}-\cite{CAT2}. 
\newline
Could $\la B \ra_W$ yield the elusive value of  $\bf B$ in the presence of interference? If so,  one can have a component of a spin-$1/2$ 
equal to $100$ \cite{SPIN100}, a negative kinetic energy, a system \e{disembodied from its properties} \cite{CAT1}, a  flux of \e{disembodied conserved properties} \cite{CAT2}, or a particle's velocity exceeding the speed of light  \cite{REL1}. The question is whether  these results provide a further insight into the 
\e{machinery behind the law} \cite{FeynL}. As always, the double slit case provides for a simple test. 
Let $\B$ measure the number of the slit chosen by the system, $B_1=1$ and $B_2=2$. Finding $\la B \ra_W=-100$, 
one must conclude that, with only two slits available, the system passes on average, through a slit number $-100$.
\newline
Could a WV be the \e{occupation number}, counting the number of particles, or particle-antiparticle pairs, as was suggested in \cite{OCCUP}? Again, in a double-slit experiment where two paths lead to a dark fringe, $\A_1=-\A_2$, the \e{occupation numbers} given 
by WV's of the projectors $\B=|b_1\ra \la b_1|$ and $\B=|b_2\ra \la b_2|$ are $\infty$ and $-\infty$, respectively. The fact that the system never arrives in its final state appears to rely on the presence (in some strange sense) of infinitely many copies of the system in each of the paths leading to it. 
\newline
Finally, and at first glance most reasonably, is simply to say that if the 
 WV of a projector is $0$ (the weak pointer has not moved), the system has definitely {\it not} taken the chosen path \cite{PHOT2}, \cite{Matz1}. Conversely, if the pointer has moved, the system must have been present in the corresponding pathway. This is, however, the same as saying that the system is present, at the same time, in all pathways where the amplitudes $\A_l$ do not vanish. But, insisted the authors of \cite{FeynL}, \cite{FeynC}, \cite{FeynH},  this would be a wrong conclusion. Besides, in the previous two-slit example, there will be be particles travelling towards a dark fringe via each of the 
slits, yet never arriving there at all. Similarly, in the three-slit example of Sects. VI and VII the system, this time arriving in its final state, 
can be \e{seen} in each of the two paths, yet conspicuously absent from their union. 
\newline
The difference between the two approaches is now evident. 
None of the pictures just discussed serve as an explanation of the interference mechanism, as long as an explanation 
is expected to let one see a new phenomenon through a prism of accepted concepts.
Rather, they look like a catalogue of illustrations of how an attempt to look inside the mechanism may end in the proverbial \e{blind alley}
of \cite{FeynC}.  
\newline 
There is, however, one more point to make. 
What could be said against using the \e{weak measurements}
to portray the failures  of classical concepts  as  novel \e{paradoxical} properties of the quantum world?
There is an inexhaustible supply of such \e{paradoxes}, which can be displayed until the reader, or the editor, gets tired. 
To take a simple example, it is an experimental fact that  in a double-slit experiment where $\alpha_1,\alpha_2 \ne 0$, the weak pointers, measuring projectors on each path, have both \e{moved}. Hence, in the quantum world, a system can be in two places at the same time  \cite{3BOX}.
In the orthodox view \cite{FeynL}, \cite{FeynC}, \cite{FeynH}, this conclusion is wrong, since an additional measurement never finds 
parts of the system distributed between the pathways. To which an advocate of \e{weak measurements} can reasonably reply that 
he/she was referring to an unobserved system, and not to the one disturbed by this additional measurement. And if it is not in both places, 
then where was it? You can't speak about this in the standard approach, replies the orthodox. And why is that? Because if I try to check your prediction, I only find it in one place. But I was referring to an unobserved system ... at which point the conversation begins to circle. 
Apparently, Feynman has thought about this, since in \cite{FeynC} he added: 
\e{ You can always {\it say} [that a particle goes through one hole or the other]- provided you stop thinking immediately and make no deductions from it. Physicists prefer not to say it, rather than stop thinking at the moment.}
The same is true of other conclusions based on the \e{weak values}, or indeed on any approach relying on the probability amplitudes, rather than on the probabilities. One can {\it say}, for example,  that the electron and its charge go different ways  \cite{CAT1},  
but the notion is so narrow that 
it is better not to say it at all. The barely speakable is, in practice, rather unspeakable. 
\newline 
Finally, it is remarkable the \e{weak measurements} approach, which purports  \e{to revisit the whole quantum mechanical notion that physical variables are described by Hermitian operators} \cite{WV2010} has not,  until now, been measured against the textbook orthodoxy of  \cite{FeynL}, \cite{FeynC}, \cite{FeynH}.
Above we have tried to do just that, and even if the reader remains unconvinced by our arguments, he/she may be interested in 
revisiting the subject and drawing his/her own conclusions. }
{
\section {Appendix A. Measurements: \e{Strong} and \e{weak},  Local and non-local.}
Suppose Bob makes three measurements of $\a$, $\B$ and $\C$.
Bob will use three von Neumann pointers \cite{vN} with positions $f_{A}$, $f_{B}$, and $f_{C}$,  prepared in the initials states 
  \begin{eqnarray}\label{B0}
|D^I(0)\ra= \int G(f_I,\Delta f_I)|f_I\ra df_I, \q I=A,B,C.
\end{eqnarray}
The functions  $G(f_I,\Delta f_I)$, whether complex or real valued, have a property [$\delta(f)$ is the Dirac delta]
  \begin{eqnarray}\label{B-1}
|G(f_I,\Delta f_I\to 0)|^2 \to \delta(f_I), 
\end{eqnarray}
and become very broad and slow-varying as $\Delta f_I \to \infty$ . The pointers, which have no own dynamics, can be briefly coupled to the system by means of  $\h^{int}_I= -i\partial _{f_{A}}\hat I \delta(t-t_I)$,  $t_I=t_0,t_1,t_2$,
so that each interaction entangles a pointer with the system.
 For example, we have $\u^{int}_B(t) \equiv \exp[\int_{t_1-\epsilon}^{t_1+\epsilon}\h^{int}_B dt]=\sum_{l=1}^N
 |b_l\ra\la b_l|\exp(-i\partial_{f_B} B_l)$, so that ($|\psi_S\ra$ is an arbitrary system's state)
 \begin{eqnarray}\label{B-2}
|\u^{int}_B(t_1)|D^B(0)\ra |\psi_S\ra =\sum_{l=1}^N \la b_l|\psi_S\ra \left [\int G(f_B-B_l, \Delta f_B)|f_B\ra df_B \right ] |b_l\ra \n
\equiv \sum_{l=1}^N
 \la b_l|\psi_S\ra |D^B(B_l)\ra|b_l\ra,
\end{eqnarray}
 and similarly for the other two pointers. 
\subsection{Unitary evolution of the composite}
At the end of his experiment Bob wants to know the likelihood of seeing particular pointer's readings $f_A$, $f_B$ and $f_C$. 
Alice is only able to make the prediction, provided the first accurate measurement,  $\Delta f_A \to 0$, prepares  the composite $\{\it system \q (S)  + three\q pointers\}$  in a known initial state
 \begin{eqnarray}\label{B-3}
|\Phi(t_0)\ra = |D^C(0)\ra|D^B(0)\ra|D^A(A_k)\ra|a_k\ra. 
\end{eqnarray}
 Note  that  the eigenvalue $A_k$ must not be degenerate. 
The composite undergoes unitary evolution until $t=t_2$, 
 $|\Phi(t_2)\ra=\u^{S+Pointers}(t_2,t_1)|\Psi(t_0)\ra$, where
  \begin{eqnarray}\label{B-4}
 \u^{S+Pointers}(t_2,t_1)= \u^{int}_C(t_2)\u_S(t_2,t_1)\u^{int}_B(t_1)\u_S(t_1,t_0)
\end{eqnarray}
and $\u_S(t',t)$ is the system's evolution operator. It is readily seen that with $\h^{int}_I$ chosen as above,
   \begin{eqnarray}\label{B-5}
|\Phi(t_2)\ra=\sum_{n=1}^N\sum_{l=1}^N \A(c_n\gets b_l \gets a_k)|D^C(C_n)\ra|D^B(B_l)\ra| D^A(A_k)\ra|c_n\ra,
\end{eqnarray}
 where $\A(c_n\gets b_l \gets a_k)\equiv \la c_n|\u^S(t_2,t_1)|b_l\ra \la b_l|)\u^S(t_1,t_0)|a_k\ra$.
 \newline
 By (a) and  (e) of Sect. III, there always exist probabilities of finding the composite in any one of the orthogonal states 
 $|f_C\ra|f_B\ra|f_A\ra|c_n\ra$, $P(f_C,f_B,f_A,c_n) =|\la f_C|\la f_B|\la f_A|\la c_n|\Phi(t_2)\ra|^2$, 
which  Alice needs to sum over the system's states $|c_n\ra$, in order to obtain the desired probability, 
\begin{eqnarray}\label{B-6}
W(f_C,f_B,f_A)=\sum_{n=1}^NP(f_C,f_B,f_A,c_n)=\q\q\q\q\q\q\q\q\q\n
\sum_{n=1}^N\left |\sum_{l=1}^N \A^S(c_n\gets b_l \gets a_k)G(f_C-C_n) G(f_B-B_l)G(f_A-A_k)\right |^2,
\end{eqnarray}
where $\C|c_n\ra =C_n|c_n\ra$, and $\int df_Adf_Bdf_C W(f_C,f_B,f_A)=1$.  
\subsection{Broken unitary evolution of the measured system}
The pointers' readings are only interesting insofar as they tell something about the measured system.
Whatever can be learnt in this manner will, however, be expressed in terms of the transition 
amplitudes $\A^S(c_n\gets b_l \gets a_k)$ corresponding to the system's unitary evolution { interrupted } each times it interacts with one of the pointers.
If the last measurement is also accurate, $\Delta f_C \to 0$, $G^*(f_C-C_{n'})G(f_C-C_{n})=\delta(f-C_n)\delta_{nn'}$,
Eq.(\ref{B-6}) reduces to 
 \begin{eqnarray}\label{B-7}
W(f_C,f_B,f_A)=
\delta (f_A-A_k) \sum_{n=1}^N\delta (f_C-C_n)\left |\sum_{l=1}^N \A^S(c_n\gets b_l \gets a_k) G(f_B-B_l)\right |^2.
\end{eqnarray}
The probability of having at $t=t_1$  a reading  $f_B$,
while measuring 
 $A_k$ and  $C_n$ at $t=t_0$ and  $t=t_2$, respectively, is, therefore,  given by
 \begin{eqnarray}\label{B-8}
W(C_n, f_B,A_k)=
\left |\sum_{l=1}^N \A^S(c_n\gets b_l \gets a_k) G(f_B-B_l)\right |^2 \equiv\left |\sum_{l=1}^N \A_l(n)G(f_B-B_l)\right |^2
\end{eqnarray}
(In the case of a degenerate eigenvalue $C_m$  one would need to sum Eq.(\ref{B-8})
over all $|c_n\ra$ corresponding to the same $C_m$.)
\newline
As an example, consider measuring  at $t=t_1$  a projector  $\B=|b_2\ra \la b_2|$ , with a non-degenerate eigenvalue $1$, and $(N-1)$-fold degenerate 
 eigenvalue $0$. 
 This is a double-slit problems with $N$ final states, sketched in Fig.5.
\begin{figure}[h]
\includegraphics[angle=0,width=12cm, height= 8cm]{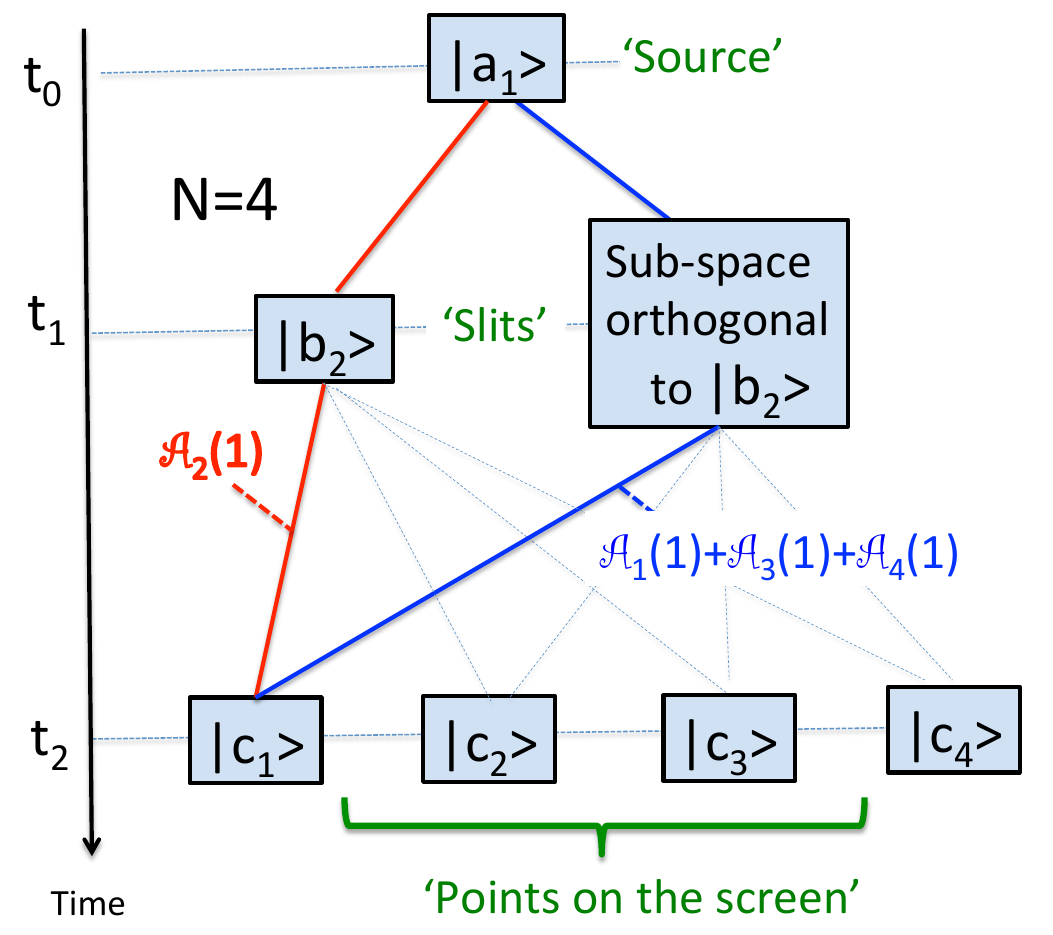}
\caption {A \e{double-slit problem} arising from measuring a projector $\B=|b_2\ra\la b_2|$
in $N=4$ dimensions.
There are two virtual routes  to a final state $|c_1\ra$, one passing through $|b_2\ra$ at $t=t_1$, 
and the other not. The corresponding amplitudes are $\A_2(1)$ 
and $\sum_{l\ne2}\A_{l}(1)$, respectively.
[$\A_l(n)\equiv \la c_n|\u^S(t_2,t_1)|b_l\ra \la b_l|\u^S(t_1,t_0)|a_1\ra$.]
By measuring  both $\B$ and $\C$ accurately, Bob can determine the route chosen by the system. }
\label{Fig.1}
\end{figure}
\subsection{Accurate \e{strong} measurements and locality}
If the second  measurement is also accurate, $\Delta f_B \to 0$, 
there are only two possible readings, $f_B=1$ and $f_B=0$,
 \begin{eqnarray}\label{B1}
W(C_n, f_B,A_k) \xrightarrow[f_B \to 0 ]{} |A_2(n)|^2\delta(f_B-1) + \left |\sum_{l\ne 2}\A_l(n)\right|^2\delta(f_B).
\end{eqnarray}
Notably the likelihood
of the system being seen to take a path $\{c_n\gets b_2 \gets a_k\}$
depends only on the corresponding amplitude, 
$\A^S(c_n\gets b_{2} \gets a_k)$. 
This can be seen as a \e{local property}. In an optical realisation of the experiment, 
 the probability of arriving at a detector  via one arm of the interferometer
would not depend on what happens elsewhere in the setup.
This dependence does, however,  appear  if one calculates the relative probabilities of arriving at the same
detector via different arms. Indeed, given that the system arrives in $|c_n\ra$, for the frequency 
with which a path $\{c_n\gets b_2 \gets a_k\}$ is chosen
one has
 \begin{eqnarray}\label{B2}
P(c_n\gets b_2 \gets a_k)=
 \frac{|\A_2(n)|^2}{|\A_2(n)|^2+\left |\sum_{l\ne 2} \A_l(n)\right |^2}. 
\end{eqnarray}
This is, however, still a \e{local} result since  two probabilities in the denominator are evaluated separately for each 
of the routes shown in Fig.5. The same can be said about the mean pointer reading, also  given by 
 \begin{eqnarray}\label{B3}
 \la f_B\ra \equiv \frac{\int f_B W(C_n, f_B,A_k) df_B }{\int W(C_n, f_B,A_k) df_B} \xrightarrow[\Delta f \to 0 ]{} \frac{|\A_2(n)|^2}{|\A_2(n)|^2+\left |\sum_{l\ne 2} \A_l(n)\right |^2}.   
\end{eqnarray}
\subsection{Inaccurate \e{weak} measurements and non-locality}
In the opposite limit, where  the measurement of $\B$ is highly inaccurate (\e{weak}), 
$ \Delta f_B \to \infty$, the interference between the paths in Fig.5 remains (almost) intact. 
Expanding  the broad $G$ in a Taylor series, $G(f_B-1,  \Delta f_B ) \approx G(f_B,  \Delta f_B )-\partial_{f_B}G(f_B,  \Delta f_B )$, and keeping
only the terms linear in the derivatives  yields 
 \begin{eqnarray}\label{B4}
W(C_n, f_B, A_1)\approx  
 |G|^2\left|\sum_{l=1}^N\A_{l}(n)\right|^2
 -\partial_{f_B}|G|^2\R\left [\A_2(n)\sum_{\l'=1}^N\A^*_{l}(n) \right ]\q\n
  -\partial_{f_B}\varphi |G|^2I\left[\A_2(n)\sum_{l=1}^N\A^*_{l}(n)\right ]\equiv W_0(C_n, f_B, A_1)+\delta W(C_n, f_B, A_1),\q
\end{eqnarray}
where $G(f_B,  \Delta f_B )=|G|\exp (i\varphi)$.
In Eq.(\ref{B4}), $W_0(C_n, f_B, A_1)$, is the probability of finding the pointer at an $f_B$ without the measurement.
The second term, $\delta W(C_n, f_B, A_1)$, is the change caused by the pointer's weak interaction with the system,
but now the \e{local property} of an accurate maesuremen is lost.
For  {any} $f_B$, $\delta W(C_n, f_B, A_1)$ depends on the amplitudes $\A_{l}(n)$ of all paths  leading to the same final state $|c_n\ra$, and not of just the one upon which $\B$ projects.
It is not possible to deduce the probability of taking the path $\{c_n\gets b_2 \gets a_k\}$ the way it was done in Eq.(\ref{B2}), 
but one can still evaluate the mean pointer's reading in Eq.(\ref{B3}). 
For a $G$ real and symmetric,  $G(f_B)=G^*(f_B)$, $G(f_B)=G(-f_B)$,
one finds
 \begin{eqnarray}\label{B5}
\la f_B \ra \xrightarrow[\Delta f \to \infty ]{} \R\left [ \frac{\A_2(n)}{\A_2(n)+\sum_{l\ne 2}\A_{l}(n)}\right ].
\end{eqnarray}
Equations (\ref{B3}) and (\ref{B5}) both involve all  path amplitudes $\A_l$, and yet are very different.
Consider again an optical version of the experiment. 
A  projector $\B=|b_2\ra\la b_2|$ is supposed to check the presence of the system in a particular arm of the interferometer,
and it does so if the measurement is accurate. To calculate, from Eq.(\ref{B1}), the probability that the pointer will move, 
one needs to know only one path amplitude, $\A_2(n)=0$, which is a local property of the system. To calculate the distribution of the readings in Eq.(\ref{B4}),  one requires the knowledge of
the amplitudes for all pathways connecting the source with the detector. Consequently, the average (\ref{B5}), computed with the  help
of such a distribution, 
is not a valid  indicator of what happens locally in a chosen pathway. 
One exception is the case where a null mean reading of a \e{weak} pointer does indicate that the $\A_2(n)=0$ vanishes, 
but that, of course, can also be verified by measuring the projector $\B$ accurately.
\newline
In summary, quantum mechanics can always  be used to compute, in a standard way,  the distribution of the pointers' 
reading, or use the distribution to evaluate the averages. For a pre- and post-selected ensemble, the results
are always expressed in terms of probability amplitudes of the pathways connecting its initial and final states. 
An accurate pointer destroys the interference, and the absolute squares of the amplitudes arise naturally as probabilities, or observed frequencies. 
A weak inaccurate pointer leaves the interference intact, and a different combinations of the amplitudes occur. 
Whether such combinations can have their  own physical meaning, 
is the main question we ask in this paper. 

{
\section {Appendix B. Some Gaussian integrals.}
Evaluating the integrals we obtain 
 \begin{eqnarray}\label{C1}
 \int\om_1(f,f')dfdf'=|\A|^2,\q\q\q\q\q\q\q\q\q\q\q\q\q\q\q\q\q\n
\int\om_2(f,f')df df' = |\A_2|^2[ 2 -2(\pi \Delta f^2/2)^{-1/2} \times \q\q\q\q\q\q\q \n
 \int \exp[-(f-1)^2/\Delta f^2]
\exp(-f^2/\Delta f^2) df]
\approx  |\A_2|^2/\Delta f^2,\q\q\q  
\end{eqnarray}
and 
\begin{eqnarray}\label{C2}
\int f \om_1(f,f') df df'=0,\q\q\q\q\q\q\q\q\q\q\q\q\q\q\q\q\q\q\q\q\q\q\q\q\q\q\q\q\n
\int f \om_2(f,f') df df' =|\A|^2[1-2(\pi \Delta f^2/2)^{-1/2}
 \int \exp[-(f-1)^2/\Delta f^2]\exp(-f^2/\Delta f^2)df)\q\n
  =|\A|^2
[1-\exp(-1/2\Delta f^2)]\approx |\A|^2/2\Delta f^2, \q\q\q\q\q\q\q\q\q\q\q\q\q\q\q\q\q\q\q\q
\end{eqnarray}
so  that
\begin{eqnarray}
\int f\rho_1(f,f')=\int f \om_1(f,f') df df'/\int\om_1(f,f')dfdf' =0,\q\q\q\q\q\q\q\q\q\q\q\q\n
\int f\rho_2(f,f')=\int f \om_2(f,f') df df'/\int\om_2(f,f')dfdf' \approx 1/2.\q\q\q\q\q\q\q\q\q\q\q
\end{eqnarray}
}

{
\section*{Acknowledgements}
DS acknowledges financial support by the Grant PID2021-126273NB-I00 funded by MICINN/AEI/10.13039/501100011033 and by \e{ERDF A way of making Europe}, as well as by the Basque Government Grant No. IT1470-22.}

\begin{thebibliography}{99}
\bibitem{FOOT1} Note that the rule (e) protects the scheme of Sect.  III in case one decides to measure, immediately after $t_2$, an operator $\hat D$, which commutes with $\C$, but has only  eigenvalues. Such a measurement would destroy interference between all paths passing through $|c_l\ra$ at $t_2$. However, causality requires that the outcomes, already obtained, cannot be affected by future measurements. The only way to avoid a contradiction is to postulate (e).
\end{thebibliography}
\end{document}